\documentclass[a4paper, groupaddress]{jpconf}
\usepackage{graphicx}
\begin{document}

\title{Project 8 detector upgrades for a tritium beta decay spectrum using cyclotron radiation}

\author{
A~Ashtari~Esfahani$^1$,
S~B{\"o}ser$^2$,
C~Claessens$^2$,
L~de~Viveiros$^3$,
P~J~Doe$^1$,
S~Doeleman$^4$,
M~Fertl$^1$,
E~C~Finn$^5$,
J~A~Formaggio$^6$,
M~Guigue$^5$,
K~M~Heeger$^7$,
A~M~Jones$^5$,
K~Kazkaz$^8$,
B~H~LaRoque$^3$,
E~Machado$^1$,
B~Monreal$^3$,
J~A~Nikkel$^7$,
N~S~Oblath$^5$,
R~G~H~Robertson$^1$,
L~J~Rosenberg$^1$,
G~Rybka$^1$,
L~Salda{\~n}a$^7$,
P~L~Slocum$^7$,
J~R~Tedeschi$^5$,
T~Th{\"u}mmler$^9$,
B~A~Vandevender$^5$,
M~Wachtendonk$^1$,
J~Weintroub$^4$,
A~Young$^4$ and
E~Zayas$^6$
}

\address{
$^1$ Center for Experimental Nuclear Physics and Astrophysics, and Department of Physics, University of Washington, Seattle, WA, USA\\
$^2$ Johannes Guttenberg University, Mainz, Germany\\
$^3$ Department of Physics, University of California, Santa Barbara, CA, USA\\
$^4$ Harvard-Smithsonian Center for Astrophysics, Cambridge, MA, USA\\
$^5$ Pacific Northwest National Laboratory, Richland, WA, USA\\
$^6$ Laboratory for Nuclear Science, Massachusetts Institute of Technology, Cambridge, MA, USA\\
$^7$ Department of Physics, Yale University, New Haven, CT, USA\\
$^8$ Lawrence Livermore National Laboratory, Livermore, CA, USA\\
$^9$ Karlsruhe Institute for Technology, Karlsruhe, Germany
}

\ead{ezayas@mit.edu}

\begin{abstract}
Following the successful observation of single conversion electrons from $^{83m}$Kr using Cyclotron Radiation Emission Spectroscopy (CRES), Project 8 is now advancing its focus toward a tritium beta decay spectrum. A tritium spectrum will be an important next step toward a direct measurement of the neutrino mass for Project 8. Here we discuss recent progress on the development and commissioning of a new gas cell for use with tritium, and outline the primary goals of the experiment for the near future.
\end{abstract}

\section{An Overview of CRES and Phase I Success}

Project 8 was conceived with the goal of directly measuring the neutrino mass from the tritium beta decay spectrum endpoint. Very near the endpoint, the spectrum will be influenced by a nonzero effective mass of the electron neutrino, $m_\beta$. Direct measurement experiments compare an empirical spectrum to the $m_\beta=0$ prediction and look for a significant distortion to extract the neutrino mass. Currently, all such measurements are consistent with $m_\beta=0$ and thus produce only limits on the neutrino mass. The strongest present limit comes from the Mainz \cite{mainz} and Troitsk \cite{troitsk} experiments: $m_\beta<2$~eV \cite{pdg}, and KATRIN anticipates a sensitivity of $m_\beta<200$~meV \cite{katrin}. In its final phase, Project 8 will target a sensitivity as low as 40~meV.

The method employed by KATRIN and the Mainz and Troitsk experiments is limited in sensitivity to about 200~meV; thus, an experiment which aims to improve on KATRIN-level sensitivity must use a different technique. Project 8 has pioneered a technique called Cyclotron Radiation Emission Spectroscopy (CRES) to measure the tritium spectrum by detecting cyclotron radiation from individual beta electrons in a magnetic field. Fourier analysis~\cite{p8} of the cyclotron signal provides a precise measurement of the frequency, which is directly related to the electron energy:
\begin{equation}
\label{1}
2\pi f=\frac{eB}{K+m_e}
\end{equation}

\noindent where $f$ is the frequency, $e$ is the elementary charge, $B$ is the (ideally uniform) magnetic field magnitude, $K$ is the electron kinetic energy and $m_e$ is the electron mass. Fig. 1 shows an event detected in the first phase of Project 8, which established CRES as a viable technique for a tritium endpoint experiment \cite{p8prl}.

\begin{figure}[!htb]
\begin{minipage}{0.5\textwidth}
\centering
\includegraphics[width=7cm]{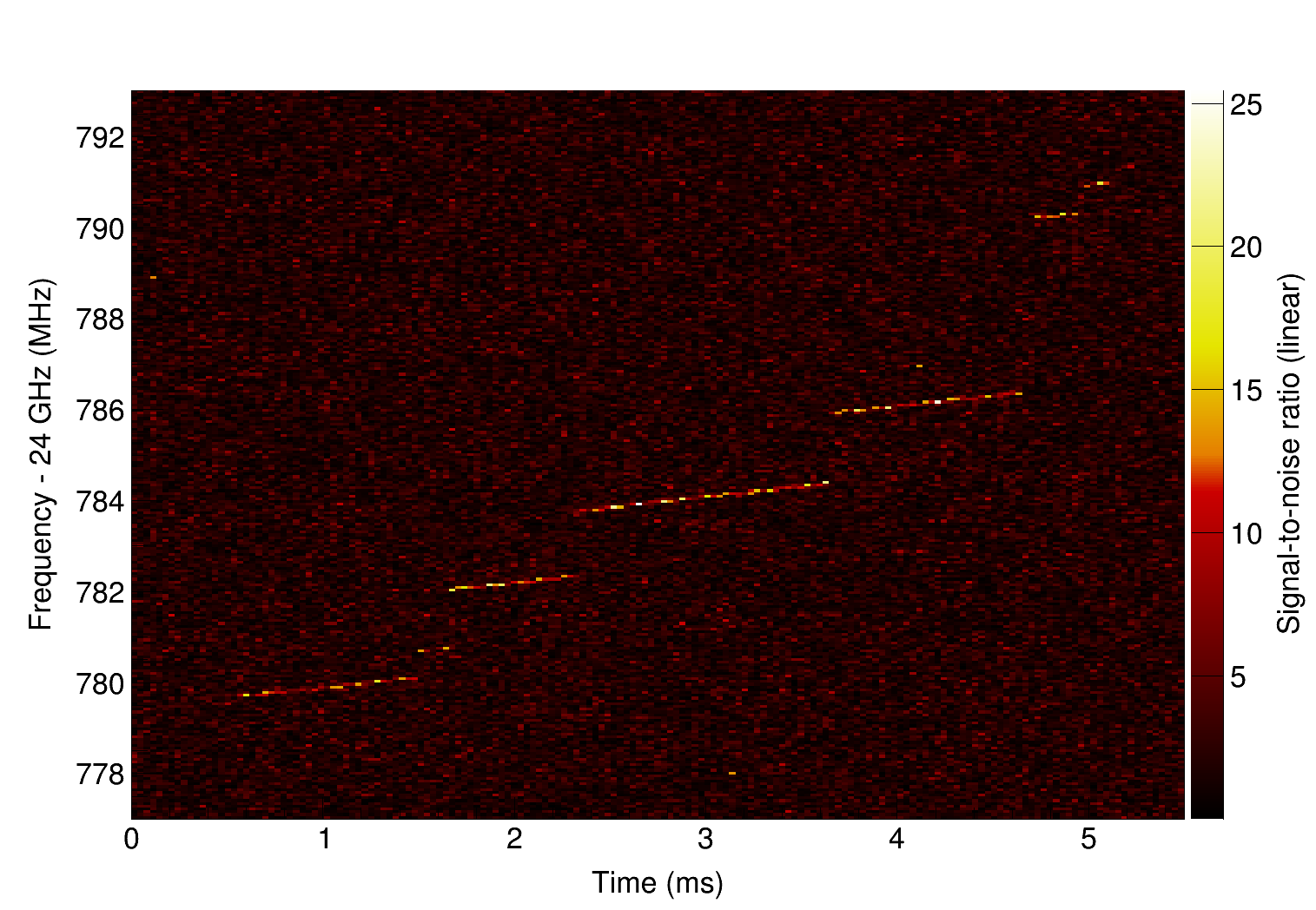}
\end{minipage}
\begin{minipage}{0.5\textwidth}
\caption{``Event 0'', the first cyclotron signal detected by Project 8 in 2014. The frequency of the signal increases with time due to radiative energy loss, and the discrete steps correspond to sudden scatters off of residual gas molecules.}
\end{minipage}
\end{figure}

\section{The Path Forward: A Tritium Spectrum}

The long-term plan of Project 8 is organized into four phases:

\begin{itemize}
\item{{\bf Phase I (completed):} proof of concept, demonstration of CRES, and measurement of the krypton conversion electron spectrum.}
\item{{\bf Phase II (current):} collection of the first ever beta decay spectrum of molecular tritium using CRES.}
\item{{\bf Phase III:} extraction of a limit competitive with the current Mainz/Troitsk limits of approximately 2~eV. Phase III will have a much larger fiducial volume than previous phases, and will collect data for approximately 1 year.}
\item{{\bf Phase IV:} first ever direct measurement of the neutrino mass. Phase IV will pioneer the use of {\it atomic} tritium to improve the sensitivity as low as 40~meV.}
\end{itemize}

\section{Phase II Cell and Waveguide Plans}

In Phase II, we will construct two new cells: one for use with krypton and one with tritium. The two cells will follow from the same design and will ideally be exact replicas. Many significant upgrades have been incorporated in the new cell design, including:

\begin{itemize}
\item{A more elaborate magnetic trap, consisting of 5 solenoidal trapping coils. This will allow us to better study the magnetic trap and related systematics of the detector.}
\item{A reliable probe of the field strength using electron spin resonance (ESR). These ESR probes are installed very close to the trap coils to ensure high accuracy, and are currently being used to calibrate the magnetic trap.}
\item{A cell volume which is a factor of 3 larger than in Phase I, with a cylindrical waveguide geometry to further increase the effective trap volume.}
\item{A reflector at one end of the cell which will coherently reflect the signal power, ideally increasing the signal-to-noise (S/N) ratio by a factor of 2.}
\item{A circulator to greatly reduce thermal noise in the waveguide to approximately 50 K, further improving the S/N ratio.}
\end{itemize}

The krypton cell, pictured in Fig. 2, was constructed in the Spring of 2016 and is now being commissioned for use with the detector. We have already observed a much lower noise floor due to the circulator, and thus expect a greatly enhanced S/N ratio over Phase I. The primary goals of commissioning this cell are to study/optimize the magnetic trap, and to directly test all of the Phase II design requirements except tritium compatibility. The tritium cell is currently under construction, and will be used to collect Project 8's first ever beta decay spectrum. We plan to begin collecting data with tritium before the end of 2016.

\begin{figure}[h!]
\centering
\includegraphics[width=13cm]{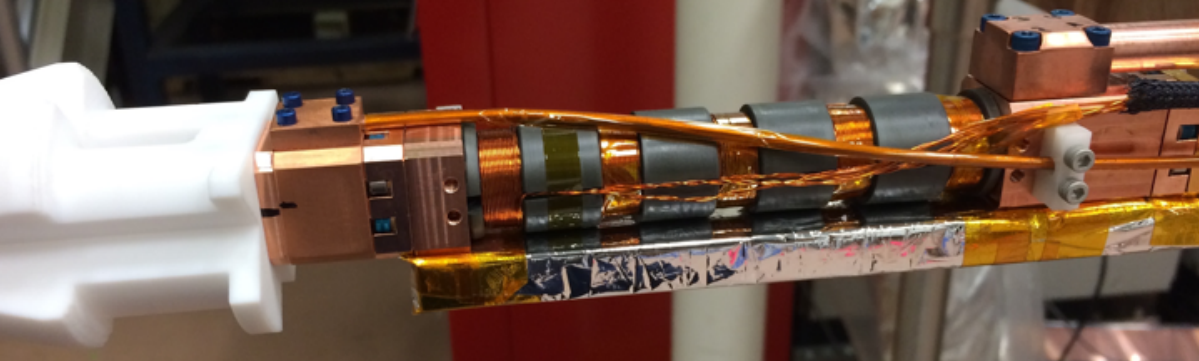}
\caption{Photograph of the krypton cell and waveguide for Phase II. The five trapping coils can be seen in the center, an extension from the 3-coil Phase I design.}
\end{figure}

\section{Signal Analysis Development}

In addition to the upgrades discussed in the previous section, we also anticipate the need for more sophisticated analysis in Phase II. The krypton spectrum consists of discrete lines, while the tritium spectrum is continuous; Phase I analysis often looks at these krypton peaks individually, and any misreconstructed events outside the peaks can be easily identified. However, with a continuous spectrum this luxury is lost; poor reconstructions can masquerade as genuine signals and must be more carefully accounted for. In addition, the noise level from the Phase II krypton cell is substantially lower and less frequency-dependent than in Phase I. Clearer signals and a continuous spectrum both motivate the need for more quantitative spectrum analysis, which will be crucial to understand the results of Phase II and future phases as well.

\section{Outlook}

A tritium spectrum will be the next important step toward a direct measurement of the neutrino mass for Project 8. In the last year, we have made substantial upgrades to the detector for use with tritium, and we have designed and built a new gas cell which has so far met all of our expectations. With the advances in both hardware and software that are currently under way, we can work toward the high level of precision that future phases of the experiment will demand.

\end{document}